# Orthoferrite with a hidden lanthanide magnetic motif: NdFeO$_3$

S. W. Lovesey[1, 2, 3]

[1] ISIS Facility, STFC, Didcot, Oxfordshire OX11 0QX, UK

[2] Diamond Light Source, Harwell Science and Innovation Campus, Didcot, Oxfordshire OX11 0DE, UK

[3] Department of Physics, Oxford University, Oxford OX1 3PU, UK

**Abstract** Scrutiny of an established monoclinic magnetic space group for NdFeO$_3$ reveals hitherto unknown properties of the orthoferrite. Future experiments using neutron and x-ray diffraction techniques can verify them. Neodymium ions possess Dirac multipoles, both time-odd (magnetic) and parity-odd (polar), that come with unique diffraction conditions. Non-magnetic polar Nd multipoles are permitted even though the monoclinic space group is centrosymmetric. Dirac multipoles are forbidden by symmetry at sites occupied by ferric ions. Available diffraction patterns have not been analysed for Dirac multipoles. Nor all permitted components of the axial dipoles and quadrupoles. In the case of neutron diffraction, magnetic quadrupoles are correlations between anapole and orbital degrees of freedom. We give conditions for the observation of Templeton-Templeton scattering of x-rays, created by angular anisotropy in the electronic charge distribution. Axial multipoles are the sole providers of dichroic signals.

## I. INTRODUCTION

In the event that NdFeO$_3$ harbours polar magnetism, it will be a unique lanthanide orthoferrite. The possibility of Kramers Nd Dirac multipoles that are magnetic and parity-odd is one outcome of the reported symmetry-informed analysis based on an established magnetic space group for the material [1]. Along with metal-oxides CuO [2, 3], α-Fe$_2$O$_3$ [4, 5] and CoTi$_2$O$_5$ [6], for example, it is a monoclinic space group with ions in acentric sites. Other orthoferrites have unusual properties, not least a ferroelectric ground state. Such is the case of GdFeO$_3$ with magneto-structural couplings associated with rotations of corner-linked FeO$_6$ octahedra, and ferroelectric polarization and magnetization controlled by magnetic and electric fields, respectively [7, 8]. Evidently, a simple orthorhombic structure is not correct for magnetic GdFeO$_3$ although non-polar Pnma is cited for the parent structure, as it is with most lanthanide orthoferrites. Multiferroism has been created by negative pressure in EuTiO$_3$ using nanocomposite films [9]. Likewise, a stabilized hexagonal TmFeO$_3$ thin film heterostructure enabled multiferroism to be artificially imposed on a naturally orthorhombic orthoferrite [10]. Magnetic properties, including exchange interactions, of a non-Kramers lanthanide orthoferrite (TbFeO$_3$) have been derived from neutron scattering, one of two experimental techniques that we consider [11].

In more detail, NdFeO$_3$ is a soft material with an orthorhombically distorted structure derived from a cubic perovskite structure [12, 13]. Weak Fe ferromagnetism is attributed to a Dzyaloshinskii-Moriya interaction, as in haematite. Spin switching, magnetization reversal and magneto-structural coupling have been observed in single crystals of NdFeO$_3$, and the structure and magnetic properties of nanoparticles have been reported [8, 14, 15, 16]. Optical properties of orthoferrites have been been investigated experimentally and theoretically [17, 18], and material characterization, structure and preparation of NdFeO$_3$ are thoroughly treated by Serna *et al*. [19].

It is suggested that NdFeO$_3$ has monoclinic magnetic symmetry at all temperatures below T$_N$ ≈ 760 K [1]. In which case, there there are many magnetic similarities between the orthoferrite and binary Fe and O in the form of haematite ($\alpha$-Fe$_2$O$_3$). Current candidates for the magnetic space group for haematite include monoclinic C2/c (magnetic crystal-class 2/m) and C2′/c′ (2′/m′) with ferric ions in sites devoid of symmetry [4, 5]. Both centrosymmetric structures permit the piezomagnetic effect, and allowed terms in the thermodynamic potential include H and HEE (H and E are magnetic and electric fields), i.e., they are compatible with ferromagnetism and a linear magnetoelectric effect is not permitted. Evidence indicating that NdFeO$_3$ complies with a polar crystal class is reviewed in Ref. [20]. Such is the case for GdFeO$_3$, for which the polar magnetic crystal class m′m′2 is mentioned [7]. Experiments proposed here for NdFeO$_3$ on the basis of a favoured centrosymmetric space group might clinch the debate for a different and polar magnetic structure.

While the monoclinic structure P2$_1$′/c′ proposed for NdFeO$_3$ belongs to the centrosymmetric magnetic crystal class 2′/m′ ferric ions displays both ferromagnetic and antiferromagnetic continuous spin reorientations of ferric ions [21]. Specifically, Fe$^{3+}$ (3d$^5$) dipole moments continuously rotate around the crystal axis [0, 1, 0] with temperature T in the region 105 K < T < 180 K [13, 22]. Moreover, Fe ions occupy two independent centrosymmetric sites. Whereas, Nd$^{3+}$ (4f$^3$) ions occupy sites devoid of symmetry, and axial and polar neodymium magnetic multipoles are permitted. Polar magnetic multipoles include a scalar, with the same discrete symmetries as Dirac's monopole, and the dipole is often referred to as an anapole [23].

In an atomic description, a monopole ⟨**S**·**R**⟩ represents magnetic charge, where **S** and **R** are spin and orbital electronic degrees of freedom, respectively, and angular brackets denote a time-average (expectation value). Such a charge contributes to the diffraction of x-rays utilizing an electric dipole - magnetic dipole event [24, 25, 26]. An anapole is the next member along in a family of electronic Dirac multipoles and it is equivalent to a dipole ⟨(**S**×**R**)⟩ and a like orbital entity [23, 27, 28]. By dint of Neumann's principle, electronic multipoles are invariant with respect to all discrete symmetries possessed by their environment. An electronic structure factor that represents a crystalline material complies with all spatial and magnetic symmetries of the material; local symmetries, and symmetries in the space group that involve translations [27, 29]. Such a structure factor defines a Bragg diffraction pattern. Motifs of Dirac multipoles therein are not occult order parameters because they diffract x-rays and neutrons, together with charge-like distributions of electrons and nuclei [3, 30, 31]. Compton scattering of x-rays is another technique with potential to observe Dirac multipoles [32, 33]. Theoretical methods that work in tandem with mentioned experimental techniques include advanced simulations of electronic structures. There are various established codes by which to derive estimates of axial and Dirac multipoles and their contributions to x-ray absorption spectra, dichroic signals and Bragg diffraction patterns [34, 35, 36]. Analytic expressions of Dirac multipoles for V and Cu ions in $V_2O_3$ and CuO, respectively, are found in Refs. [37, 38], and axial multipoles for $Np^{4+}$ ($5f^3$) are treated in ref. [39].

The magnetic structure $P2_1'/c'$ for $NdFeO_3$ [1, 20, 21] is discussed in the following section. It is the basis of calculated Bragg diffraction patterns presented in Sections III, IV and V. Sections III and IV deal with neodymium and ferric ions, respectively. Each section is subdivided into results for neutron diffraction [12, 13, 40, 41, 42] and resonant x-ray diffraction [24, 27, 43, 44, 45]. A Bragg diffraction pattern is composed of core spots derived from neutron diffraction by nuclei or x-ray diffraction by spherical distributions of electronic charge. The core data define a chemical structure. Core data are overlayed by weak, or timid, or basis-forbidden, Bragg spots that arise from angular anisotropy in the distribution of electronic charge, and referred as Templeton-Templeton (T&T) scattering [46, 47]. And magnetic Bragg spots not indexed on the chemical structure, which are the principal interest in the present study. One such finding is neutron and resonant x-ray diffraction by components of Nd and Fe axial dipoles parallel to the unique axis of the monoclinic cell. The components in question are not shown in Fig. 1, which instead depicts the currently accepted configuration of axial dipoles [1, 12, 13]. There are Bragg spots unique to the motif of Nd Dirac multipoles. No such Bragg spots are permitted for ferric ions because they occupy sites with a centre of inversion symmetry. Uniquely Nd Dirac Bragg spots appear in diffraction patterns gathered by magnetic neutron diffraction and resonant x-ray diffraction. Our calculations for x-ray diffraction incorporate the effect of rotation of the crystal about the reflection vector, because the azimuthal-angle scans yield valuable information about magnetic structure. Selection of the primary x-ray energy in resonant diffraction selects contributions from different elements, which are Nd and Fe in our examples [43, 45]. Section V describes non-magnetic polar resonant x-ray diffraction by Nd ions, which is permitted even though $P2_1'/c'$ is centrosymmetric [27, 44, 49].

## II. MAGNETIC STRUCTURE

The parent structure of the lanthanide-iron perovskite $NdFeO_3$ is taken to be non-standard orthorhombic Pbnm. Cell edges a ≈ 5.45102 Å, b ≈ 5.58808 Å and c ≈ 7.76165 Å [13]. The magnetic space group $P2_1'/c'$ (BNS setting, No. 14.79, magnetic crystal class $2'/m'$) accounts for magnetization studies and neutron powder diffraction data [1]. Axial dipoles in the standard Pnma setting are depicted in Fig. 1 [13]. Basis vectors for $P2_1'/c'$ relative to Pbnm are {(−1, 0, 0), (0, −1, 0), (1, 0, 1)} with unit cell parameters a, b and $c_o = \sqrt{[a^2 + c^2]} \approx 9.48456$ Å, and an obtuse angle = $\cos^{-1}(-a/c_o) \approx 125.08°$. Refinement of the Bragg pattern excluded neodymium Dirac multipoles, and magnetic contributions were deduced by subtraction of patterns collected above and below the onset of long-range magnetic order [12]. Greater sensitivity to the distribution of magnetization can be achieved with diffraction by a single crystal, and neutron polarization analysis [40, 42, 50, 51]. Or the diffraction of x-rays tuned to an atomic resonance, which is a subject of calculations reported herein [27, 43, 44, 45]. Reflections ($h$, 0, $l$) with $l = 2n + 1$, (0, $k$, 0) with $k = 2n + 1$ and (0, 0, $l$) with $l = 2n + 1$ are absent in $P2_1/c$. Magnetic allowed reflections that have been observed are the principal evidence in favour of $P2_1'/c'$ [12, 13, 20]. Therefore, intensities predicted for systematic absence conditions in the diffraction pattern of $P2_1'/c'$ are critical tests of its suitability for $NdFeO_3$.

The reciprocal lattice of P2$_1$′/c′ has a volume v$_o$ = (abc), and vectors **a**\* = (2π/v$_o$) (−bc, 0, ab), **b**\* = (2π/v$_o$) (0, −ac, 0) and **c**\* = (2π/v$_o$) (0, 0, ab). The reflection vector for Bragg diffraction = (2π) (−h/a, −k/b, (h + l)/c) with integer Miller indices (h, k, l). Local axes (ξ, η, ζ) are derived from orthogonal vectors **a**\*, (0, −b, 0) and (a, 0, c). Axes (x, y, z) in Fig. 2 for x-ray diffraction and (ξ, η, ζ) coincide in the nominal setting of the crystal in an azimuthal angle scan. The principal symmetry operation in the space group is an anti-dyad parallel to the unique η axis (2$_η$′). Specifically, axial dipoles in the plane normal to the η axis are permitted for neodymium and ferric ions, as in Fig. 1, while Nd anapoles in the same plane are revealed in basis-forbidden Bragg spots (0, 0, l) with odd l. However, Nd axial dipoles parallel to the η axis are not forbidden since Nd ions occupy sites with no symmetry, cf. Eq. (8). Likewise for Fe ions that occupy centrosymmetric sites with no additional symmetry constraints, cf. Eqs. (11) and (13).

Atomic spherical multipoles ⟨O$^K_Q$⟩, that have integer rank K and projections Q in an interval − K ≤ Q ≤ K, encapsulate electronic degrees of freedom [27, 41]. For future use, 2$_η$ ⟨O$^K_Q$⟩ = (−1)$^{K + Q}$ ⟨O$^K_{−Q}$⟩. Cartesian and spherical components of a dipole **R** = (x, y, z) are related by x = (R$_{−1}$ − R$_{+1}$)/√2, y = i(R$_{−1}$ + R$_{+1}$)/√2, z = R$_0$. The complex conjugate is defined as ⟨O$^K_Q$⟩\* = (−1)$^Q$ ⟨O$^K_{−Q}$⟩, with a phase convention ⟨O$^K_Q$⟩ = [⟨O$^K_Q$⟩′ + i⟨O$^K_Q$⟩″] for real and imaginary parts labelled by single and double primes, respectively. Generic multipoles ⟨O$^K_Q$⟩ are later replaced by specific forms for neutron and x-ray diffraction.

Diffraction amplitudes presented here are specific to position multiplicity, Wyckoff letter and symmetry [27, 41, 43, 45]. Multipoles that enter diffraction amplitudes are denoted ⟨T$^K_Q$⟩ (⟨G$^K_Q$⟩) in general discussions of axial (polar) contributions with no specific type of radiation in mind. Axial multipoles are ⟨T$^K_Q$⟩ and ⟨t$^K_Q$⟩ for neutron (§§ III.A and IV.A) and resonant x-ray (§§ III.B, IV.B and V) diffraction, respectively. For the latter type of diffraction, Dirac multipoles are ⟨g$^K_Q$⟩, and polar charge-like multipoles are ⟨u$^K_Q$⟩ [27]. The Dirac dipole in neutron scattering is ⟨**D**⟩, with ⟨**H**$^2$⟩ the quadrupole of immediate interest. Ref. [41] contains a review of all Dirac multipoles in the neutron scattering amplitude.

### III. NEODYMIUM IONS

Neodymium ions (Nd$^{3+}$, $^4$I, 4f$^3$) use sites 4e in P2$_1$′/c′ that have no symmetry. Such acentric sites permit Dirac atomic multipoles that change sign with respect to individual reversals of time and space. Time and parity signatures are denoted σ$_θ$ and σ$_π$, respectively; magnetic multipoles are time odd σ$_θ$ = −1, and σ$_θ$ σ$_π$ = −1 (+1) for axial (Dirac) multipoles.

An electronic structure factor Ψ$^K_Q$ = [exp(i**κ**·**d**) ⟨O$^K_Q$⟩$_d$] where **κ** is the reflection vector, and the implied sum is over the four Nd ions in a unit cell [27, 41]. The significant result for general sites 4e in space group P2$_1$′/c′ (No. 14.79 [21]; note OG settings used in [1, 20]) is,

Ψ$^K_Q$(4e) = ⟨O$^K_Q$⟩ [αβγ + σ$_π$ (αβγ)\*]

  + σ$_θ$ (−1)$^{K + Q}$ (−1)$^{k + l}$ ⟨O$^K_{−Q}$⟩ [α\*βγ\* + σ$_π$ (α\*βγ\*)\*].          (1)

Spatial phase factors are $\alpha = \exp(i2\pi h x)$, $\beta = \exp(i2\pi k y)$ and $\gamma = \exp(i\pi l/2)$, with general coordinates $x \approx 0.2613$ and $y \approx -0.048$ [13]. Axial Nd multipoles $\langle T^K_Q \rangle$ alone contribute to bulk ferromagnetism ($h = k = l = 0$), and dipoles (K = 1) are $\langle T^1\xi \rangle$ and $\langle T^1_0 \rangle = \langle T^1\zeta \rangle$ depicted in Fig. 1.

The structure factor $\Psi^K_Q(\text{Nd})$ evaluated for basis-forbidden $(0, k, 0)$ with odd $k$,

$$\Psi^K_Q(4e) = [\beta + \sigma_\pi \beta^*] [\langle O^K_Q \rangle + (-1)^{K+Q} \langle O^K_{-Q} \rangle], \quad (0, k, 0) \text{ odd } k, \quad (2)$$

reveals axial dipoles and anapoles parallel to the $\eta$-axis. Diagonal (Q = 0) components of $\langle O^K_Q \rangle$ that can be observed have an even rank. Not surprisingly, basis-forbidden reflections $(0, 0, l)$ with odd $l$ are more revealing. Indeed, the corresponding structure factor,

$$\Psi^K_Q(4e) = [\gamma + \sigma_\pi \gamma^*] [\langle O^K_Q \rangle + \sigma_\pi (-1)^{K+Q} \langle O^K_{-Q} \rangle], \quad (0, 0, l) \text{ odd } l \quad (3)$$

is different from zero for Dirac multipoles alone. Anapoles are $\langle G^1\xi \rangle$ and $\langle G^1\zeta \rangle$, for example. An array of Dirac multipoles in a compound is an occult order parameter for many experimental investigations, with magnetic neutron and resonant x-ray diffraction notable exceptions. Lastly, spatial phases and multipoles in $\Psi^K_Q(4e)$ also factor for $(h, 0, l)$ with odd $l$. The reflection $(1, 0, l)$ is notable because $\sin(2\pi x) \approx 1.0$, which gives a strong contribution to $\Psi^K_Q(4e)$ from $\langle T^1\eta \rangle$ and a negligible contribution from Dirac multipoles.

### A. Neutron diffraction

Multipoles in neutron diffraction depend on the magnitude of the reflection vector $\kappa$ [40, 41, 42]. Radial integrals $\langle j_c(\kappa) \rangle$ in axial multipoles $\langle T^K_Q \rangle$ are averages of spherical Bessel functions and integer $c$ is even [52]. By definition, $\langle j_c(0) \rangle = 0$ for $c > 0$, and $\langle j_0(0) \rangle = 1$. Most simple models of magnetic neutron scattering are based on the dipole $\langle T^1 \rangle$. A useful approximation in terms of the rare-earth magnetic moment $\langle \mu \rangle = g \langle J \rangle$ is,

$$\langle T^1 \rangle \approx (\langle J \rangle/3) \{ \langle j_0(\kappa) \rangle g_S + [\langle j_0(\kappa) \rangle + \langle j_2(\kappa) \rangle] g_L \}, \quad (4)$$

where $g_S = 2(g - 1)$ and $g_L = (2 - g)$ [40]. The Landé factor $g = 4/3$ for $Nd^{3+}$ with $J = 9/2$. An equivalent operator $[(\mathbf{S} \times \mathbf{R}) \mathbf{R}]$ for the quadrupole $\langle T^2 \rangle$ shows that it measures the correlation between a spin anapole $(\mathbf{S} \times \mathbf{R})$ and orbital degrees of freedom [41].

The Dirac dipole $\langle \mathbf{D} \rangle$ in neutron diffraction is the sum of three contributions that include expectation values of spin and orbital anapoles. Operators for the three contributions to $\mathbf{D}$ are a spin anapole $(\mathbf{S} \times \mathbf{R})$, orbital anapole $\mathbf{\Omega} = [\mathbf{L} \times \mathbf{R} - \mathbf{R} \times \mathbf{L}]$, and $(i\mathbf{R})$. Specifically, $\langle \mathbf{D} \rangle = (1/2) [3 (h_1) \langle (\mathbf{S} \times \mathbf{R}) \rangle - (j_0) \langle \mathbf{\Omega} \rangle + (g_1) \langle (i\mathbf{R}) \rangle]$. Form factors $(h_1)$, $(j_0)$ and $(g_1)$ have been calculated for several atomic configurations [38, 53]. In what follows, we retain $\langle \mathbf{D} \rangle$ and a quadrupole $\langle \mathbf{H}^2 \rangle$ that possess the largest atomic form factors. A quadrupole of this type is a product of $(h_1)$ and a correlation function $\langle \{\mathbf{S} \otimes \mathbf{R}\}^2 \rangle$ written in terms of a standard tensor product of two dipoles [41]. Notably, $\langle \mathbf{H}^2 \rangle \propto [(h_1) \langle \{\mathbf{S} \otimes \mathbf{R}\}^2 \rangle]$ accounts for magnetic neutron diffraction by the pseudo-gap phase of ceramic superconductors [41, 54, 55].

Intensity of a magnetic Bragg spot = $|\langle\mathbf{Q}_\perp\rangle|^2$ when the neutron beam is unpolarized [40, 41]. Here, $\langle\mathbf{Q}_\perp\rangle = \{\mathbf{e} \times (\langle\mathbf{Q}\rangle \times \mathbf{e})\}$ and $\mathbf{e}$ is a unit vector in the direction of the reflection vector. The axial intermediate amplitude $\langle\mathbf{Q}\rangle^{(+)} = \langle\boldsymbol{\mu}\rangle/2$ in the forward direction of scattering ($\kappa = 0$), and the superscript $^{(+)}$ denotes parity-even, $\sigma_\pi = +1$. In general, a polarized neutron diffraction signal $\Delta = \{\mathbf{P}\cdot\langle\mathbf{Q}_\perp\rangle\}$, where $\mathbf{P}$ is polarization of the primary neutrons [50, 51]. A spin-flip intensity SF is a measure the magnetic content of a Bragg spot, and SF = $\{|\langle\mathbf{Q}_\perp\rangle|^2 - \Delta^2\}$ when $\mathbf{P}\cdot\mathbf{P} = 1$ and $(\langle\mathbf{Q}_\perp\rangle^* \times \langle\mathbf{Q}_\perp\rangle) = 0$ [56].

Reflections (0, 0, $l$) with odd $l$ possess an amplitude $\langle\mathbf{Q}_\perp\rangle^{(-)} = (0, \langle Q_{\perp\eta}\rangle^{(-)}, 0)$, $\mathbf{e} = (\cos(\chi), 0, \sin(\chi))$ with $\cos(\chi) = [\text{sign}(l)\,(a/c_o)]$ and,

$$\langle Q_{\perp\eta}\rangle^{(-)} \approx 4\sin(\pi l/2)\,\{[\cos(\chi)\langle D_\zeta\rangle - \sin(\chi)\langle D_\xi\rangle]$$
$$+ (3/\sqrt{5})[\sin(\chi)\langle H^2_{+1}\rangle'' - \cos(\chi)\langle H^2_{+2}\rangle']\}. \quad (0, 0, l) \text{ odd } l \qquad (5)$$

The quantity $[\cos(\chi)\langle D_\zeta\rangle - \sin(\chi)\langle D_\xi\rangle]$ is the component of the Nd anapole normal to the reflection vector. The monoclinic obtuse angle = $\cos^{-1}(-a/c_o) \approx 125.08°$ and $\chi = 54.92°$ for positive $l$. Reversing the sign of the Miller index $l$ reverses the sign of $\cos(\chi)$ while $\sin(\chi)$ is unchanged. The spin-flip intensity is zero for neutron polarization normal to the reflection vector. Neutron amplitudes for a centrosymmetric structure are purely real. Eq. (2) for (0, $k$, 0) with odd $k$ shows that axial and polar amplitudes are proportional to $\cos(2\pi k y)$ and $\sin(2\pi k y)$, respectively. However, the axial amplitude $\langle\mathbf{Q}_\perp\rangle^{(+)} = 0$, because $\langle\mathbf{Q}\rangle^{(+)}$ is parallel to $\mathbf{e}$. Likewise, $\langle\mathbf{Q}_\perp\rangle^{(-)} = 0$.

### B. Resonant x-ray diffraction

Rotational anisotropy in multipoles is most pronounced in the direct vicinity of an absorption edge whereas it is negligible far from the edges. Sum-rules for dichroic signals arise from quantum numbers of the core state embedded in multipoles [27, 37, 57, 58, 59]. They can be calculated from a ground state wavefunction for dichroism and scattering using standard tools of atomic spectroscopy; see Refs. [37, 38, 39] for worked examples. Tuning the energy of x-rays to an atomic resonance has two obvious benefits [43, 45]. In the first place, there is an enhancement of Bragg spot intensities and, secondly, spots are element specific. There are four scattering amplitudes labelled by photon polarization states depicted in Fig. 2, e.g., unrotated ($\sigma'\sigma$) and rotated ($\pi'\sigma$) scattering amplitudes. Strong Thomson scattering, by spherically symmetric atomic charge, that overwhelms weak signals is absent in rotated channels of polarization [44]. It is allowed in unrotated channels of polarization using a parity-even absorption, but absent in a parity-odd absorption, e.g., electric dipole - electric quadrupole (E1-E2) events exploited to observe Dirac multipoles. Diffraction amplitudes presented here include rotation of the crystal about the reflection vector (azimuthal-angle scan).

The range of values of the rank K of multipoles is fixed by the triangle rule, and K = 0 - 2 and K = 1 - 3 for a E1-E1 and E1-E2 events, respectively. For parity-even multipoles observed with an E1-E1 event the time signature $\sigma_\theta = (-1)^K$. Scattering amplitudes are proportional to radial integrals. A dipole radial integral $(\Theta|R|\Xi)$ accompanies E1 in a scattering amplitude, for example, where Θ is a valence state that carries orbital angular momentum $l$ and Ξ is a core state that carries angular momentum $l_c$ and the two angular momenta differ by unity (parity-odd). Dimensionless quantities $\Re$(E1-E1) = $[(\Theta|R|\Xi)/a_o]^2$, $\Re$(E1-E2) = $[q(\Theta|R|\Xi)(\Theta'|R^2|\Xi)/a_o^2]$ and $\Re$(E2-E2) = $[q(\Theta'|R^2|\Xi)/a_o]^2$, with $l + l'$ odd, are useful measures of scattering amplitudes, where q is the photon wavevector and $a_o$ the Bohr radius [43, 45].

Neodymium L edges have energies $L_2 \approx 6.724$ keV (E1, 2p - 5d) and $L_3 \approx 6.212$ keV. Calculations of x-ray absorption spectra and resonant diffraction intensities of intermetallic NdMg show E2 (2p - 4f) contributions below the Nd absorption edges [60]. With **b**\* and the η-axis antiparallel at the start of an azimuthal angle scan, and an E1-E2 event [44],

$$(\sigma'\sigma) \approx i \sin(\pi l/2) \cos(\theta) \sin(\psi) \{\cos(\chi) [- \langle g^1_\zeta \rangle + (1/3)\sqrt{10} \langle g^2_{+2} \rangle''] $$

$$+ \sin(\chi) [\langle g^1_\xi \rangle - (1/3)\sqrt{10} \langle g^2_{+1} \rangle'']. \quad (0, 0, l) \text{ odd } l \quad (6)$$

A complete $(\sigma'\sigma)$ for $(0, 0, l)$ odd $l$ contains octupoles, and it is relegated to an Appendix. On retaining anapoles and quadrupoles, as in Eq. (6), the rotated amplitude is [44],

$$(\pi'\sigma) \approx i \sin(\pi l/2) \cos(\theta) \cos(\psi) \{3\sqrt{10} \sin(\theta) [\cos(\chi) \langle g^1_\zeta \rangle - \sin(\chi) \langle g^1_\xi \rangle ]$$

$$- [\cos(\theta) \sin(\psi) \cos(\chi) + 2 \sin(\theta) \sin(\chi)] \langle g^2_{+1} \rangle''$$

$$- [\cos(\theta) \sin(\psi) \sin(\chi) - 2 \sin(\theta) \cos(\chi)] \langle g^2_{+2} \rangle''\}. \quad (0, 0, l) \text{ odd } l \quad (7)$$

Unlike the approximation for $(\sigma'\sigma)$, the companion $(\pi'\sigma)$ contains two harmonics of the azimuthal angle, namely, $\cos(\psi)$ and $\sin(2\psi)$. Amplitudes $(\pi'\sigma)$ and $(\sigma'\pi)$ are related by a change in the sign of the Bragg angle, θ.

Diffraction at space-group forbidden $(0, k, 0)$ is very different from the foregoing reflections, in part because now the reflection vector coincides with an axis of crystal rotation symmetry, namely, the unique η-axis. Amplitudes in unrotated channels of polarization are identically zero, $(\sigma'\sigma) = (\pi'\pi) = 0$ for $(0, k, 0)$ with odd $k$, while the rotated channel can be different from zero and displays two-fold symmetry in ψ. First, though, consider parity-even diffraction using an E1-E1 absorption event.

Templeton-Templeton scattering is created by charge-like quadrupoles $\langle t^2_Q \rangle$ [27, 46, 47]. In the present case, T&T scattering is not present in unrotated channels of polarization. The magnetic dipole $\langle t^1_\eta \rangle$ permitted in $(\pi'\sigma)$ with enhancement by an E1-E1 event is independent of the azimuthal angle, because the dipole is parallel to the reflection vector. The E1-E1 rotated channel of polarization contains an amplitude,

$$(\pi'\sigma) = (\sigma'\pi) = 4 \cos(2\pi k y) \{(i/\sqrt{2}) \sin(\theta) \langle t^1_\eta \rangle$$

$$+ \cos(\theta) [\cos(\psi) \langle t^2_{+1} \rangle'' + \sin(\psi) \langle t^2_{+2} \rangle'']\}, \quad (0, k, 0) \text{ odd } k \quad (8)$$

with **a**\* in the plane of scattering for ψ = 0. Magnetic and T&T contributions differ in phase by 90°, and intensities are in quadrature, i.e., T&T intensity is a two-fold periodic function of the azimuthal angle. Miller indices $k = 1, 3$ and 5 satisfy the Bragg condition for diffraction at the Nd L-edges. However, $\cos(2\pi k y) \approx 0$ for $k = 5$. An amplitude at the Nd M-edge (E1, 2d - 4f) is $\approx 10^3$ larger than an L-edge amplitude, and the Bragg spot (0, 1, 0) is accessible at the $M_3$ edge ($\approx 1.266$ keV) [43, 45].

Returning to diffraction enhanced by a parity-odd E1-E2 event that exposes Nd Dirac multipoles $\langle g^K_Q \rangle$, we find $(\sigma'\sigma) = (\pi'\pi) = 0$ for (0, $k$, 0) with odd $k$. Anapoles (K = 1) are absent in the rotated channels of polarization $(\pi'\sigma) = (\sigma'\pi)$ and the amplitude is two-fold periodic in the azimuthal angle,

$$(\pi'\sigma) = (i/\sqrt{5}) \sin(2\pi k y) \{[\Phi + \cos^2(\theta) \cos(2\psi)] \langle g^2_0 \rangle + 2\sqrt{(2/3)} \cos^2(\theta) \sin(2\psi) \langle g^2_{+1} \rangle'$$

$$+ \sqrt{(2/3)} [3\Phi - \cos^2(\theta) \cos(2\psi)] \langle g^2_{+2} \rangle' + 2\sqrt{(10/3)} \cos^2(\theta) [-\sin(2\psi) \langle g^3_{+1} \rangle''$$

$$+ 2\sqrt{(2/5)} \cos(2\psi) \langle g^3_{+2} \rangle'' + \sqrt{(3/5)} \sin(2\psi) \langle g^3_{+3} \rangle'']\}, \quad (0, k, 0) \text{ odd } k \quad (9)$$

with **a*** is in the plane of scattering for $\psi = 0$, and $\Phi = [3 \cos^2(\theta) - 2]$. At the Nd $L_2$ edge, $\cos^2(\theta) \approx 0.973$ and $\Phi \approx 0.918$ for $k = 1$.

### IV. FERRIC IONS

Sites 2b and 2c used by Fe ions are centres of inversion symmetry. In consequence, all electronic multipoles are of the axial type. Electronic structure factors for the two independent sites differ by a spatial phase factor. Moreover, structure factors are similar to that for Nd ions in terms of content, apart from inversion excluding Dirac multipoles. Specifically,

$$\Psi^K_Q(2b) = (-1)^{h+l} \Psi^K_Q(2c) = (-1)^h [\langle O^K_Q \rangle + \sigma_\theta (-1)^{K+Q} (-1)^{k+l} \langle O^K_{-Q} \rangle]. \quad (10)$$

Evidently, the symmetry of bulk magnetization of Fe ions and axial Nd ions is identical.

#### A. Neutron diffraction

We continue to focus on basis-forbidden reflections. There is no magnetic neutron diffraction at (0, $k$, 0) with odd $k$, because $\langle \mathbf{Q} \rangle^{(+)}$ is parallel to the reflection vector. Reflections (0, 0, $l$) with odd $l$ possess an amplitude $\langle \mathbf{Q}_\perp \rangle^{(+)} = (0, \langle Q_\eta \rangle^{(+)}, 0)$ with $\mathbf{e} = (\cos(\chi), 0, \sin(\chi))$. Retaining dipoles and quadrupoles,

$$\langle Q_\eta \rangle^{(+)} \approx 3 \langle T^1_\eta \rangle + \sqrt{3} \{\sin(2\chi) [\langle T^2_{+2} \rangle' - \sqrt{(3/2)} \langle T^2_0 \rangle] + 2 \cos(2\chi) \langle T^2_{+1} \rangle''\}. \quad (11)$$

An approximation to the transition-metal dipole,

$$\langle \mathbf{T}^1 \rangle \approx (\langle \boldsymbol{\mu} \rangle/3) [\langle j_0(\kappa) \rangle + \langle j_2(\kappa) \rangle (g-2)/g], \quad (12)$$

is often used [40]. Here, the magnetic moment $\langle \boldsymbol{\mu} \rangle = g \langle \mathbf{S} \rangle$ and the orbital moment $\langle \mathbf{L} \rangle = (g - 2) \langle \mathbf{S} \rangle$. Values of $\langle j_0(\kappa) \rangle$ and $\langle j_2(\kappa) \rangle$ for ferric ions are displayed in Fig. 2 of Ref. [61]. The quadrupole $\langle \mathbf{T}^2 \rangle$ is proportional to $\langle j_2(\kappa) \rangle$, and a non-zero value relies on an admixture of at least two values of J in the ground state [41, 62].

#### B. Resonant x-ray diffraction

Iron L edges have energies $L_2 \approx 0.721$ keV and $L_3 \approx 0.708$ keV, and the wavelengths are too large to satisfy Bragg conditions for reflections (0, $k$, 0) and (0, 0, $l$). The Fe K-edge occurs at ≈ 7.115 keV and the wavelength $\lambda \approx 1.743$ Å allows resonance-enhanced Bragg diffraction at the cited reflections. Previous studies using resonant x-ray diffraction at the K-edge of 3d-transition metal compounds include $V_2O_3$, $\alpha$-$Fe_2O_3$ and NiO [30, 45, 63, 64].

An E2-E2 event (1s - 3d) has null unrotated amplitudes for (0, $k$, 0) with odd $k$. For the rotated channels of polarization [44],

$$(\pi'\sigma) = (\sigma'\pi) \approx (1/2\sqrt{5}) \{ - i\sqrt{2} \sin(3\theta) \langle t^1_\eta \rangle + i\sqrt{(3/2)} \sin(\theta) [\Phi + 5 \cos^2(\theta) \cos(2\psi)] \langle t^3_{+1} \rangle''$$

$$- 2\sqrt{(15/7)} \cos(3\theta) [\cos(\psi) \langle t^2_{+1} \rangle'' + \sin(\psi) \langle t^2_{+2} \rangle''] \}. \quad (0, k, 0) \text{ odd } k \quad (13)$$

Hexadecapoles $\langle t^4_Q \rangle$ are omitted from Eq. (13) for brevity [44]. As might be expected, E2-E2 and E1-E1 amplitudes are very similar. Replacement of $\theta$ in Eq. (8) by $3\theta$ in Eq. (13) is one main difference, together with magnetic octupoles in Eq. (13). The amplitude $(\sigma'\sigma)$ can be different from zero for (0, 0, $l$) with odd $l$, unlike space-group forbidden (0, $k$, 0). A notable feature of $(\sigma'\sigma)$ for an E2-E2 event is that it contains the dipole moment, which is not so for an E1-E1 event. Retaining multipoles with ranks K = 1, 2 and 3, as in Eq. (13), we find,

$$(\sigma'\sigma) \approx i \sin(2\theta) \sin(\psi) \{ - \sqrt{(1/10)} \langle t^1_\eta \rangle + \sqrt{(3/10)} [1 - 5 \cos^2(\chi) \cos^2(\psi)] \langle t^3_{+1} \rangle''$$

$$- \sqrt{3} \sin(2\chi) \cos^2(\psi) \langle t^3_{+2} \rangle'' + (1/\sqrt{2}) [1 + (3 \cos^2(\chi) - 4) \cos^2(\psi)] \langle t^3_{+3} \rangle'' \}$$

$$- \sqrt{(3/7)} \sin^2(\theta) \sin(2\psi) [\cos(\chi) \langle t^2_{+1} \rangle'' + \sin(\chi) \langle t^2_{+2} \rangle'']. \quad (0, 0, l) \text{ odd } l \quad (14)$$

As before in Eq. (6), **b**\* and the $\eta$-axis are antiparallel for $\psi = 0$. The magnetic part of $(\sigma'\sigma)$ is a linear combination of odd harmonics $\sin(\psi)$ and $\sin(3\psi)$, while charge-like quadrupoles are proportional to $\sin(2\psi)$.

## V. POLAR MULTIPOLES

Although the magnetic space group $P2_1'/c'$ belongs to a non-polar crystal class ($2'/m'$) polar Nd multipoles $\langle u^K_Q \rangle$ are permitted [27, 44]. They contribute with K = 1, 2 and 3 to diffraction patterns enhanced by an E1-E2 event. Dipoles $\langle \mathbf{u}^1 \rangle$ behave as displacements with regard to symmetry. A dipole $\langle u^1_\eta \rangle$ occurs in $(\pi'\sigma)$ for (0, 0, $l$) with odd $l$. There are no dipoles in $(\sigma'\sigma)$ and $(\pi'\pi)$. The latter contain quadrupoles $\langle u^2_Q \rangle'$ and octupoles $\langle u^3_Q \rangle''$, and the amplitudes are proportional to $\sin(2\psi)$ and odd functions of the Bragg angle $\theta$. Diffraction at (0, $k$, 0) with odd $k$ is simpler, with $(\sigma'\sigma) = (\pi'\pi) = 0$. Dipoles $\langle u^1_\xi \rangle$ and $\langle u^1_\zeta \rangle$ contribute to $(\pi'\sigma) = (\sigma'\pi)$, which is proportional to $\sin(2\theta)$ and a sum of $\sin(\psi)$ and $\cos(\psi)$ alone, e.g., [$\sin(\psi) \langle u^1_\xi \rangle$] and [$\cos(\psi) \langle u^1_\zeta \rangle$].

## VI. CONCLUSIONS

A survey of magnetic properties of the lanthanide orthoferrite $NdFeO_3$ has revealed significant unanswered questions. Specifically, we pursued ramifications of a magnetic space group for the material using calculated Bragg diffraction patterns [1, 20]. Our results point to future experiments that can reveal hitherto unknown properties that make $NdFeO_3$ a unique orthoferrite. Likewise, future simulations of the electronic structure can develop our understanding of its magnetic properties.

In common with several other metal-oxides, the favoured magnetic space group for $NdFeO_3$ is monoclinic (orthoferrites possess an orthorhombic chemical structure). Findings include a motif of neodymium Dirac multipoles that are parity-odd (polar) and time-odd (magnetic). The motif is visible in the diffraction of neutrons, and x-rays tuned in energy to an Nd atomic resonance. Anapoles (Dirac dipoles) are confined to the a-c plane that accommodates axial (conventional) magnetic dipoles. Fortunately, reflection conditions for Nd Dirac multipoles are unique, meaning that axial multipoles are forbidden in diffraction under the conditions required for Dirac multipoles. Our predicted amplitudes include Dirac quadrupoles and octupoles in addition to anapoles, and x-ray amplitudes include rotation of the $NdFeO_3$ crystal about the reflection vector (an azimuthal-angle scan). To date, axial dipoles parallel to the unique axis of the monoclinic unit cell have not been observed in diffraction experiments [12, 13]. We provide the reflection conditions, and those for Templeton-Templeton scattering caused by angular anisotropy in the Nd charge distribution [46, 47]. Non-magnetic polar Nd multipoles are permitted even though the monoclinic space group is centrosymmetric, and relevant x-ray diffraction amplitudes are discussed in Section V.

Ferric ions occupy sites in the monoclinic unit cell that are centres of inversion symmetry, and it forbids Dirac multipoles. Diffraction by Fe axial dipoles is the main source of data for the assignment of the monoclinic space group that we have scrutinized [12, 13]. No data have been reported for dipoles parallel to the unique monoclinic axis, and we delineate conditions for future diffraction experiments. The amplitude for magnetic neutron scattering includes a quadrupole that can exist if the Fe ground state uses two or more J-states [41, 62]. An equivalent quadrupole operator is a product of the spin anapole and orbital operator, whose existence in the ground state reveals currently unknown properties.

Dichroic signals are obtained from structure factors Eq. (1) and Eq. (10) for Nd and Fe ions, respectively, evaluated with Miller indices $h = k = l = 0$. Parity-odd signals are forbidden for both ion types. Linear dichroism using electric dipole (E1) and electric quadrupole (E2) absorption events is created by real parts of quadrupoles and hexadecapoles with projections ±2, namely, $\langle t^2_{+2}\rangle'$ and $\langle t^4_{+2}\rangle'$ [24, 27]. Magnetic circular dichroism from and E1 event is created by $\langle t^1\zeta\rangle$, the component of the axial dipole parallel to the local $\zeta$-axis in the a-c plane of the chemical structure, cf. Fig. 2 [24, 27, 65]. An E2 absorption event reveals the same dipole and the diagonal component of the octupole $\langle t^3_0\rangle$.

**Acknowledgements** D. D. Khalyavin and K. S. Knight gave much advice on crystal physics and Bragg diffraction.

### APPENDIX

With the reflection vector (0, 0, $l$) and $-\xi$ aligned, as in Fig. 2, the electronic structure factor becomes $(-1)^Q d^K_{Qq}(\chi) \Psi^K_q$, with an implied sum on projections q. The argument $\chi$ of the Wigner rotation matrix satisfies $\cos(\chi) = [\text{sign}(l) (a/c_o)]$. For the Nd reflection (0, 0, $l$) with odd $l$, an E1-E2 absorption event, and abbreviations $C = \cos(\chi)$, $S = \sin(\chi)$,

$$(\sigma'\sigma) = (4i/5)\ \sqrt{3}\ \sin(\pi l/2)\ \cos(\theta)\ \sin(\psi)\ \{C\ [-\ \langle g^1_\zeta\rangle + (1/3)\sqrt{10}\ \langle g^2_{+2}\rangle'']$$

$$+ S\ [\langle g^1_\xi\rangle - (1/3)\sqrt{10}\ \langle g^2_{+1}\rangle''] + \sqrt{(2/3)}\ C\ [1 - 5\ C^2\ \cos^2(\psi)]\ \langle g^3_0\rangle$$

$$+ (1/3)\sqrt{2}\ S\ [1 - 15\ C^2\ \cos^2(\psi)]\ \langle g^3_{+1}\rangle' + (2/3)\ \sqrt{5}\ C\ [1 - 3\ (1 + S^2)\ \cos^2(\psi)]\ \langle g^3_{+2}\rangle'$$

$$-\sqrt{(10/3)}\ S\ [C^2 + (1 - 4\ C^2)\ \cos^2(\psi)]\ \langle g^3_{+3}\rangle'. \quad (A1)$$

Anapole and quadrupole contributions feature in Eq. (6), and numerical coefficients in Eq. (A1) comply with Ref. [44]. Evidently, octupole contributions to $(\sigma'\sigma)$ modify the simple $\sin(\psi)$ azimuthal-angle dependence that hallmarks anapoles and quadrupoles. The reciprocal lattice vector **b**\* and the η-axis are antiparallel at the start of an azimuthal angle scan, where $(\sigma'\sigma) = 0$.

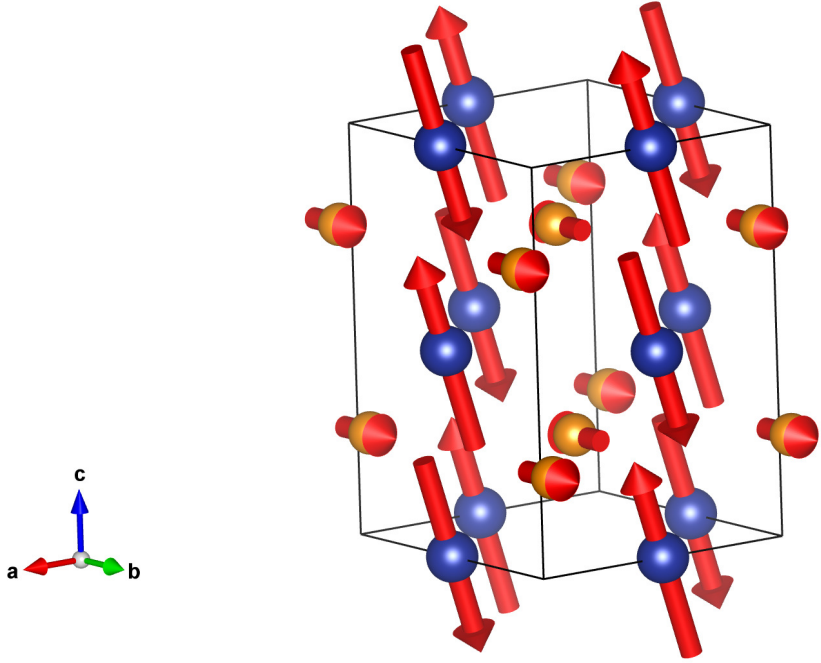

FIG. 1. Nd and Fe axial dipoles in the plane normal to the η-axis depicted as yellow and blue in the standard setting Pnma(a, b, c) ≡ Pbnm($b_o$, $c_o$, $a_o$) [13]. Reproduced from MAGNDATA [48].

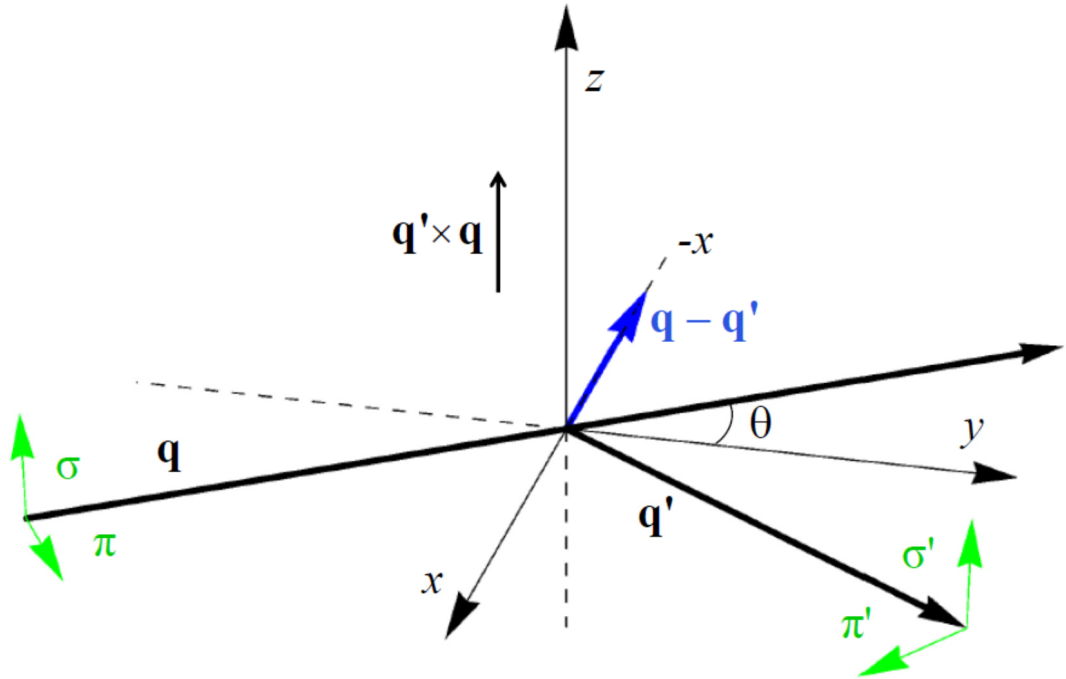

FIG. 2. Primary (σ, π) and secondary (σ′, π′) states of polarization. Corresponding wave vectors **q** and **q′** subtend an angle 2θ. The Bragg condition for diffraction is met when **q** − **q′** coincides with the reflection vector indexed (h, k, l). Orthogonal vectors **a***, (0, −b, 0) and (a, 0, c) that define (ξ, η, ζ) and depicted Cartesian (x, y, z) coincide in the nominal setting of the crystal